# Thermomechanical properties of zirconium tungstate / hydrogenated nitrile butadiene rubber (HNBR) composites for low-temperature applications

Anton G. Akulichev[a], Ben Alcock[b], Avinash Tiwari[a] and Andreas T. Echtermeyer[a]

[a] Norwegian University of Science and Technology, NTNU, Trondheim, Norway

[b] SINTEF Materials and Chemistry, Oslo, Norway

Abstract

Rubber compounds for pressure sealing application typically have inferior dimensional stability with temperature fluctuations compared with their steel counterparts. This effect may result in seal leakage failures when subjected to decreases in temperature. Composites of hydrogenated nitrile butadiene rubber (HNBR) and zirconium tungstate as a negative thermal expansion filler were prepared in order to control the thermal expansivity of the material. The amount of zirconium tungstate ($ZrW_2O_8$) was varied in the range of 0 to about 40 vol. %. The coefficient of thermal expansion (CTE), bulk modulus, uniaxial extension and compression set properties were measured. The CTE of the $ZrW_2O_8$ filled HNBR decreases with the filler content and it is reduced by a factor of 2 at the highest filler concentration used. The filler effect on CTE is found to be stronger when HNBR is below the glass transition temperature. The experimental thermal expansion data of the composites are compared with the theoretical estimates and predictions given by FEA. The effect of $ZrW_2O_8$ on the mechanical characteristics and compression set of these materials is also discussed.

# 1 Introduction

Many components of engineering equipment, such as those used in oil and gas production systems, contain elastomeric seals whose performance might be impaired at low temperatures in freezing regions or during winter seasons. The duration of individual cold cycles may not be very extensive, but they can be extremely detrimental for the pressure integrity of such seals. One of the renowned problems with elastomeric seals exposed to subzero temperatures is their considerable thermal contraction with cooling, as the coefficient of thermal expansion (CTE) of rubbers above the glass transition temperature is at least an order of magnitude greater than that of steel [1-4]. Even without taking into account other factors, such as stress relaxation, thermal contractions alone , may be responsible for a drop of the seal contact pressure by up to 50% [5]. This reduction in contact pressure can in turn lead to the formation of leak paths for contained fluids and potentially catastrophic seal failure. In this respect, measures to alleviate the dimensional change of the elastomeric seals by the development or selection of a suitable material and/or adequate seal and its housing design become of paramount importance for pressure containing systems subjected to continuous or periodic freezing. This paper deals with the former approach by reporting composite materials based on hydrogenated nitrile butadiene rubber (HNBR).

In order to compensate for the large thermal expansivity of a polymer material, it is rather natural to integrate another material having a much lower CTE into it, to reduce the overall expansivity of the composite [6]. The exploitation of materials exhibiting the negative thermal expansion (NTE) phemomenon could combat the shrinkage of the elastomer during cooling. Materials with negative thermal expansions have been known and explored for more than five decades [7] and progress both in understanding the NTE mechanisms and development of new compounds has been reported over the last twenty years [8,9]. Newly developed materials with a strong NTE effect, e.g. nitrides $Mn_3(Cu_xSi_yGe_{1-x-y})N$ [10] and $(Mn_{3+x}Zn_ySn_{1-x-y})N$ [11] or bismuth nickelates [12,13], demonstrate a substantial reduction of CTE in blends with polymers. Authors of the latter work were able to successfully produce an epoxy compound with 18 vol.% filler exhibiting zero thermal expansion in a temperature range above the ambient. Another promising NTE compound family is $A_2M_3O_{12}$ (in which A could be Al, Sc, Fe, In, Ga, or Y, and M could be Mo, or W [14]). $Al_2Mo_3O_{12}$ was recently applied to control the

CTE of polyethylene [15] and yielded a substantial suppresion of the expansivity at low filler loadings.

In the work presented in this paper, zirconium tungstate, $ZrW_2O_8$, has been selected as a filler for an HNBR elastomer as it is one of the most studied NTE materials with proven CTE of approximately $-9x10^{-6}$ $K^{-1}$ ($\alpha$-phase) [16] in the temperature range relevant for the sealing application. The material has been previously investigated in application with some polymeric materials such as polyester [17], epoxies [17-20], polyimide [21-23], polycarbonate [24], phenolic resin [25] yielding a substantial reduction of thermal expansivity of these polymers. Nanoparticles of zirconium tungstate have also recently been reported to show a CTE as low as $-11x10^{-6}$ $K^{-1}$ [26].

As such, the research reported in academic literature has so far concentrated on embedding of zirconium tungstate (or other NTE compounds) into rigid polymers and no works dedicated to softer polymers like rubber with NTE fillers have been identified. A theoretical study [27] was recently published, providing analytical estimates of the effect of NTE material in elastomer and elastomer seals, thus predicting the potential performance of this material combination.

Hence, the main objective of this research article is to report the effect of zirconium tungstate as a NTE filler on the thermomechanical behaviour (with the emphasis on the CTE) of a model rubber compound. The second objective is to find which CTE prediction model can give the most accurate property estimates for elastomers filled with NTE inclusions for potential material and product design. The third objective is to check the impact of the NTE phase on the compression set property which is a key indicator of the potential sealing function of a material.

## 2 Experimental details
### 2.1 Materials and processing

The properties of the composites were measured on specimens made from an unfilled HNBR and HNBR filled with various concentrations of zirconium tungstate. A zirconium tungstate powder was obtained for the experiments from Alfa Aesar. The powder has the particle size distribution (PSD) depicted in Figure 1. The particle size was measured using a Malvern G3 Particle size analyser based on optical analysis of images of dispersed particles at objective magnifications of 50, 10 and 2.5.

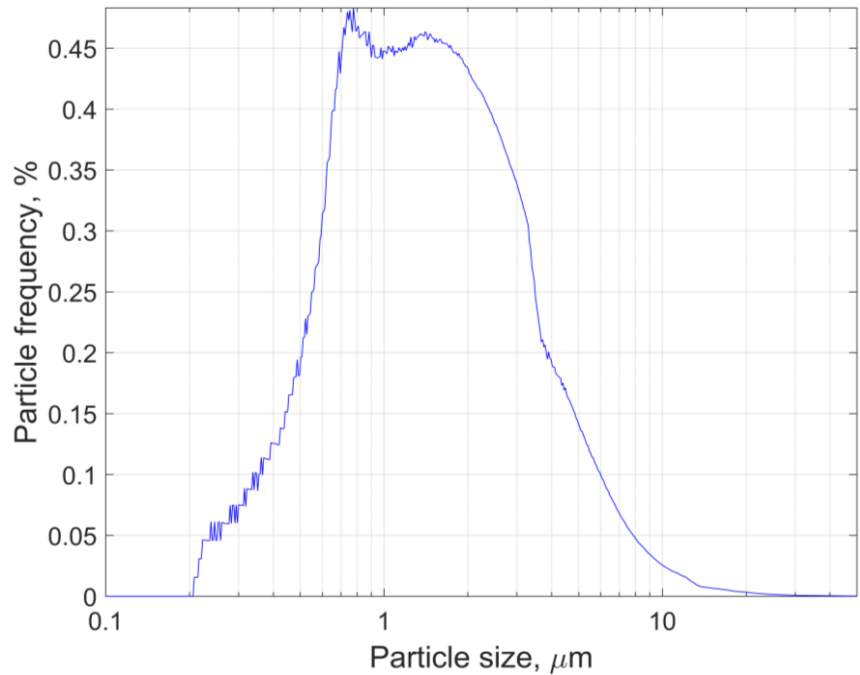

Figure 1. Zirconium tungstate particle size distribution

The filler particles vary in size from less than 0.3 μm to about 100 μm, with the majority being in the range of about 0.5 to 4 μm. Approximately 8 % of the particles have the size of less than 0.5 μm and about 1 % are larger than 10 μm.

The rubber used in the work reported here is a rubber formulation typical of that which might be used in the oil and gas industry in sealing application. The composition is shown in Table 1 and is based on hydrogenated nitrile butadiene rubber (HNBR) with 96 % saturated butadiene units with 36% acrylonitrile content. This HNBR was selected because previous work showed that it has a combination of good ageing resistance, hydrocarbon resistance [28], necessary barrier property [28,29] and some low-temperature flexibility. The compound formulation did not contain carbon black, in order to elucidate the pure effect of zirconium tungstate without having interaction effects coming from carbon black; only necessary additives and processing aids were present. Since the amount of the additives is small as compared to the added $ZrW_2O_8$, they are disregarded in the subsequent simulation and data analysis.

Table 1    Composition of the generic HNBR used in this study

| Component | Content, parts per hundred rubber |
|---|---|
| HNBR | 100 |
| Antioxidant | 3 |
| Stearic acid | 0,5 |
| Zinc oxide | 5 |
| Magnesium oxide | 10 |
| Plasticizer | 20 |
| Peroxide | 10 |

The compound described in Table 1 (except the peroxide) was combined in an internal mixer to yield a single HNBR masterbatch which was used for subsequent production of all of the materials presented in this paper. This HNBR masterbatch was then combined with $ZrW_2O_8$ and the peroxide using a Schwabenthan Polymix 110P open two roll mill. Each formulation was subjected to continious mixing on the mill for 10 minutes. All together 6 formulations with an increasing content of zirconium tungstate of 0, 8.6 %, 17.3 %, 25.3 %, 35.8 %, 39.7 % volume fraction were obtained.

After compounding, the materials were compression moulded into 2 mm thick sheets and test-specific specimens using a Fontijne TB200 hot press. The materials were cured at 443 K (170 °C) for 20 min in the press, followed by post-curing at 423 K (150 °C) for 4 h in an oven.

## 2.2    Characterization

2.2.1    Scanning Electron Microscopy

A Quanta FEG 650 scanning electron microscope (SEM) operated in a low-vacuum mode at 80 Pa chamber pressure and 5 kV accelerating voltage was employed to observe the microstructure of the composites at various magnifications and qualitatively evaluate the distribution of the $ZrW_2O_8$ filler particles. The fractured specimens from tensile tests (as discussed later) were examined in the microscope. No specific surface treatment was applied to the specimens.

2.2.2  Dilatometry

Thermal expansion measurements were carried out on a Netzsch DIL402C dilatometer during heating from 193 K to 473 K (-80 to 200 °C) at a heating rate of 2 K per minute. The specimens for dilatometry were 6x6x10 mm moulded-to-shape bricks. The linear CTE, $\alpha$, is computed by linear regression of the thermal dilatation data in the ranges from 193 to 233 K and from 298 to 403 K for the glassy and rubbery regions of HNBR respectively.

2.2.3  Differential Scanning Calorimetry (DSC)

The glass transition temperature, $T_g$, of the cured rubber compound was determined by DSC using a Perkin Elmer DSC 8500 at a heating rate of 20 K per minute and appeared to be 250 K (-23 °C). The DSC plot is given in Figure 2.

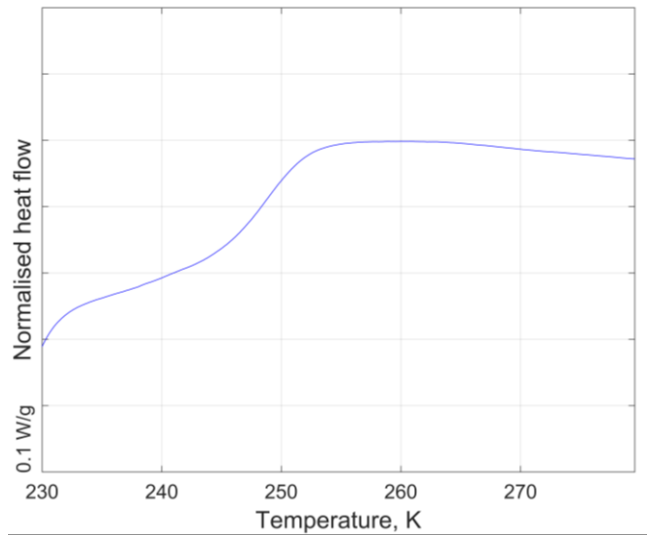

Figure 2. DSC curve of the HNBR compound used in this study; heat flow is normalized per unit mass of specimen, oriented with endotherm up.  The glass transition of the material is clearly visible as a step in the curve, and $T_g$ reported in this paper is determined by the mid-point of the transition.

2.2.4  Thermogravimetric Analysis (TGA)

The actual mass and the corresponding volume fraction of $ZrW_2O_8$ in the prepared formulations were determined on Perkin Elmer Pyris 1 TGA machine by burning the rubber matrix and measuring the mass of the residuals. The tests were performed in accordance with the procedure described in ISO 9924-3 – procedure A. A sample of 5 mg was heated in nitrogen at 20 K/min from 296 K to 873 K, cooled in nitrogen from 873 K to 673 K, held for 2 minutes in air at 673 K, heated from 673 K to 923 K at 20 K/min and held in air at 923 K for 10 minutes. The mass of residuals obtained in each compound constituted the mass of $ZrW_2O_8$ and other non-combustible components in HNBR, which was measured separately through burning the virgin rubber compound and subtracted from the total mass of residuals. The volume fractions of the zirconium tungstate in the compounds are then calculated using the density of zirconium tungstate. One sample was normally taken from each compound for the TGA, however an estimate of the sample variance was also obtained by means of two replicas in the compound with the largest property variance. The standard deviation of mass was found as low as 0.29 % .

2.2.5  Bulk Modulus Measurement

The bulk modulus was measured on a special test set-up for volumetric compression measurements that comprises a steel pressure vessel and a pump system with a Quizix C-5000 pump cylinder, a set of hoses and two valves. The test stand is schematically depicted in Figure 3. A specimen was inserted into the pressure vessel which was then fully filled with distilled water and connected to the pump by a short metallic tube. The pump cylinder provides the system with high pressure monitored by a pressure gauge. The measurement is performed by tracking the change of external pressure in the system against the amount of hydraulic fluid (distilled water) supplied to the vessel when the delivery valve is open. The water supply rate was 3 ml/min.

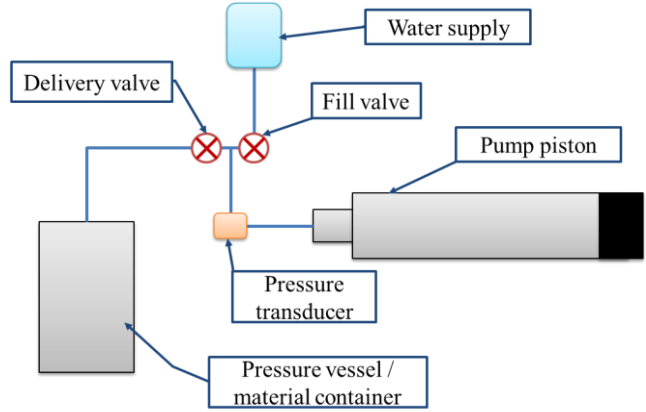

Figure 3. Schematic of the volumetric compression test set-up

The outcome of the measurement is given as a pressure versus supplied volume curve. Only the linear part of the curve is used in the calculation of the bulk modulus via the well-known expression

$$K = V \frac{dP}{dV}. \tag{1}$$

Where $K$ is the bulk modulus; $P$ is the external pressure and $V$ is the compressed volume. Water compressibility and any expansion of the equipment are also taken into account by making a separate measurement in the system solely filled with water. Separation of the specimen compressibility and the contribution arising from the system can be done using the effective bulk modulus $K_{eff}$ obtained in the measurement. The final equation for calculating the specimen bulk modulus is

$$K_s = \frac{V_s}{\left( \dfrac{V_{eff}}{K_{eff}} - \dfrac{(V_{eff} - V_s)}{K_{sys}} \right)}. \tag{2}$$

Where $V_{eff}$ is the effective (total) volume inside the pressure vessel which includes the specimen volume $V_s$, $K_s$ is the sought specimen bulk modulus and $K_{sys}$ is the system expansion modulus computed from the abovementioned water compression experimental data.

As the bulk modulus is directly proportional to the specimen volume used, the volumetric compression tests were performed on formulations for which sufficient sample mass was available for 3 tests. The accuracy of a similar rig for measuring compressibility has been previously evaluated in detail [30] and appeared comparable with the restrained compression method. As compared to the apparatus reported in reference [30], the accuracy of the volume change measurement has been further improved by a high-precision piston position sensor. Therefore the major portion of the observed variation is believed to come from the specimen variability. All measurements were carried out at ambient temperature.

2.2.6    Uniaxial Tensile Testing

The mechanical behaviour in uniaxial extension was investigated using a Zwick universal testing machine equipped with a contact extensometer and a 2.5 kN load cell. Five test specimens were stamped out from one moulded sheet of each compound. The specimen geometry was in accordance with ISO 37 type 2. Each specimen was fixed in the machine by mechanical grips and pre-loaded to 0.5 N before stretching. All extension experiments were carried out until specimen failure at a strain rate of 0.083/sec and under ambient conditions.

2.2.7    Compression Set Measurement

Measurement of the compression set was performed at room temperature using a special mechanical fixture with the design similar to the one recommended by ISO 815-1. Cylindrical specimens of 20 mm diameter and 10 mm height were placed between thick steel plates of the fixture and mechanically compressed by 4 bolts to 80 % of the initial height defined by inserted steel spacers. The test methodology followed ISO 815-1 and the compression set values were calculated in accordance with the standard.

**2.3 Finite element analysis (FEA)**

FEA is used in this work to predict the thermomechanical properties of the composites in question (with particular focus on CTE) based on a-priori knowledge of the properties of the constituents. Abaqus software for 3D numerical simulation of the HNBR-$ZrW_2O_8$ composites was used.

A numerical homogenization technique [31,32], in which a representative volume element (RVE) of the composite is first subjected to mechanical strain to deduce effective stiffness matrix and then to a temperature change to derive CTE, is used here. The RVE is taken as a cube-shaped element with spherical filler particles regularly packed in hexagonal arrangement (the face centered cubic unit cell). SEM revealed that the real filler particles were not exactly spherical, as described in the experimental part, but this simple choice gave reasonably good modelling results. Due to symmetry, 1/8 fraction of the unit cell is sufficient for the study [32]; the RVE is depicted in Figure 4. The filler volume fraction is controlled by the particle radius. Ideal filler – matrix interface (no debonding) is implied in this approach.

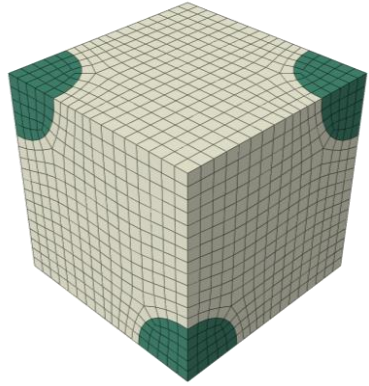

Figure 4. Meshed RVE

The homogenized composite material is assumed to obey the Hooke's law. Even though the matrix is actually non-linear hyperelastic, this assumption is reasonable since small deformations are considered. From the thermoelasticity equation the CTE can then be computed via

$$\bar{\boldsymbol{\alpha}} = \frac{(\bar{\boldsymbol{\varepsilon}} - \bar{\mathbf{S}} : \bar{\boldsymbol{\sigma}})}{\Delta T} \quad , \tag{3}$$

where $\bar{\boldsymbol{\sigma}}$ is the 2$^{nd}$ rank effective stress tensor; $\bar{\boldsymbol{\varepsilon}}$ is the 2$^{nd}$ rank effective elastic strain tensor; $\bar{\mathbf{C}}$ is the 4$^{th}$ rank effective stiffness tensor; $\bar{\mathbf{S}} = \bar{\mathbf{C}}^{-1}$ is the effective compliance tensor; $\bar{\boldsymbol{\alpha}}$ is the 2$^{nd}$ rank effective thermal expansion tensor and $\Delta T$ is the temperature change. The stiffness tensor has only two independent components in the case of isotropic material. In order to find them the RVE is stretched in one direction to 0.1 % nominal strain with all normal displacements set to 0, except, of course, one in the stretching direction. Temperature is kept constant i.e. $\Delta T = 0$. Use is made of the sum of nodal reaction forces on relevant cube faces to calculate the effective stress tensor components and further stiffness tensor components. After that, the CTE is calculated using (3) from the effective thermal stress acting in the RVE due to temperature change $\Delta T$. The effective strain tensor components in (3) are zero, since the applied boundary conditions constrains all normal displacements.

Table 2 lists material properties and element types used in the FEA. The model was meshed with quadratic brick elements. The material parameters for HNBR were obtained in respective testing except for the low-temperature data. The Neo-hookean form [33] of the strain energy density function was chosen as a rubber material model for the FEA since only moderate strains are of concern in the current investigation. The strain energy density function in this model for a compressible material is typically formed of the deviatoric and the volumetric terms [34,35] and expressed as:

$$W = C_1(\bar{I}_1 - 1) + D_1(J - 1)^2, \tag{4}$$

where $C_1$ and $D_1$ are material parameters reflecting resistance to shear and compressibility of the material respectively; $\bar{I}_1 = J^{-2/3} I_1$ is the the first deviatoric strain invariant; $J = \lambda_1 \lambda_2 \lambda_3$ and $I_1 = \lambda_1^2 + \lambda_2^2 + \lambda_3^2$ is the first strain invariant expressed in terms of the principal stretches $\lambda_i$. In the case of small strains (<1 %) the model matches Hooke's law, and the parameter $C_1$ can be related to the Young's modulus, $E$, via the known expression: $E \approx 6C_1$.

The material model was fitted to the uniaxial tensile test data in the range of extension of 1.005-1.05 using *nlfit* function from MATLAB Statistics toolbox to obtain the material parameter $C_1$. Parameter $D_1$ can be found from the bulk modulus through the following expression:

$$D_1 = \frac{2}{K}. \tag{5}$$

Table 2  Material properties used in the FEA.

| Attribute | ZrW$_2$O$_8$ Filler | HNBR Matrix | | |
|---|---|---|---|---|
| Model | Linear elastic | Linear elastic – rubbery state | Neo-hookean hyperelastic | Linear elastic – glassy state |
| Material constants | $E$ = 88000 MPa [36]<br>$v$ = 0.3 [36]<br>$a$ = -9 x 10$^{-6}$ 1/K [16] | $E$ = 6.6 MPa<br>$v$ = 0.49 and 0.4995<br>$a$ = 187 x 10$^{-6}$ 1/K | $C_1$ = 1.1 MPa<br>$D_1$ = 0.001 MPa$^{-1}$<br>$a$ = 187 x 10$^{-6}$ 1/K | $E$ = 2500 MPa<br>$v$ = 0.33<br>$a$ = 80 x 10$^{-6}$ 1/K |

| Type of elements | C3D20R | C3D20R (C3D20RH) | C3D20RH | C3D20R |
|---|---|---|---|---|

Due to lack of low-temperature test data, the HNBR Poisson's ratio, $v$, below $T_g$ is taken as 0.33, which is common for polymer glasses [37]. This choice of $v$ implies the equality between the Young's modulus and the bulk modulus.

## 3. Results and discussion
### 3.1 Scanning Electron Microscopy

The microstructure of the obtained composites, as illustrated in Figure 5, consists of mostly the HNBR elastomer (dark grey) and zirconium tungstate particles (light). The distribution of the $ZrW_2O_8$ particles appears to be uniform, although the variation of the particle sizes is substantial. The size ranges seen here appear to be from fractions of 1 micron to several tens of microns. Furthermore, the filler particles are not spherical, but slightly angular and some tend to be more preferentially elongated than others. This shape factor may somewhat affect the analytical and numerical property predictions based on perfect spheroid form of the inclusions.

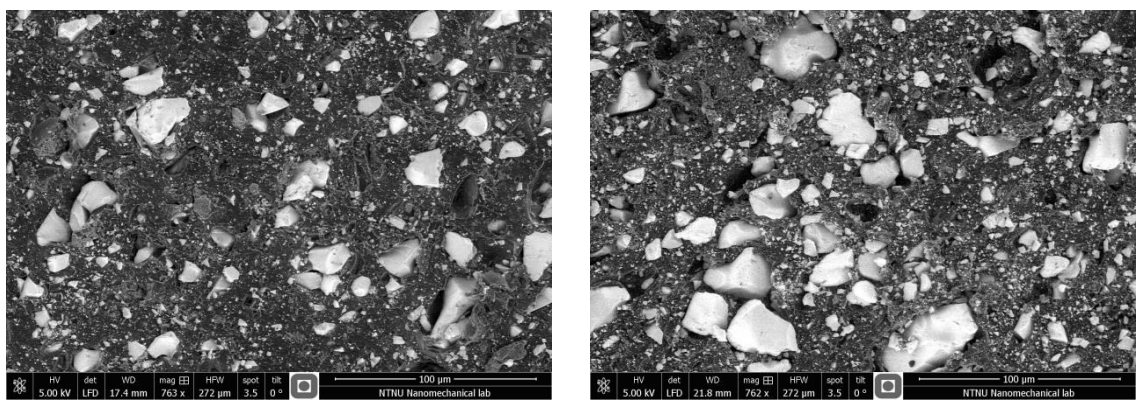

a)          b)

Figure 5. SEM images of fracture surface of HNBR-$ZrW_2O_8$.
a) 25.3 vol. % of the $ZrW_2O_8$ and b) 39.7 vol. % of the $ZrW_2O_8$

It is also noticeable that separation occurs at the filler-matrix interface, probably due to the action of high strains in the tensile test. Most black looking spots in the photographs are due to detached filler particles in the fracture zone, which is also evident from the straight edges resembling the filler inclusions. The optimisation of the $ZrW_2O_8$-HNBR interface was not of primary interest in this work, and was not investigated further in this work.

**3.2 Coefficient of Thermal Expansion**

Figure 6 illustrates the experimentally determined relationship between the CTE of the composite material in the rubbery state and $ZrW_2O_8$ volume fraction, between 298 K and 403 K, i.e. above the glass transition of the HNBR. As expected, the CTE of the HNBR-$ZrW_2O_8$ composites has a downward trend if plotted against the $ZrW_2O_8$ volume fraction. The CTE of the most filled compound (ca. 40% volume fraction) is 54% of the value measured in the original HNBR. Thereby, a potential drop of the contact pressure in a seal exposed to freezing can be reduced by a factor of 2 in this case which is in agreement with the results obtained previously [27]. It is necessary to note the CTE measurements were performed using specimens produced under laboratory conditions and not subjected to multiple thermal cycling or weathering as would be expected in the real applications. The application environment would be imperative to consider when further developing this technology to determine of the CTE of the composites is constant over multiple thermal cycles.

Figure 5 also shows predictions given by FEA, the Rule of Mixtures (ROM), Kerner model [38] and Schapery model [39] with Hashin-Shtrikman bounds [40] for the effective bulk modulus. As pointed by Schapery [39], the upper bound matches with Kerner's expression. The model equations used are compiled in Table 3 for clarity.

Table 3.     Used analytical expressions for composite CTE prediction

| Model | Expression |
|---|---|
| ROM | $\bar{\alpha} = V_f \alpha_f + (1 - V_f) \alpha_m$ |
| Kerner [38] | $\bar{\alpha} = V_f \alpha_f + (1 - V_f) \alpha_m + \dfrac{\left(V_f (1 - V_f)(\alpha_f - \alpha_m)(K_f - K_m)\right)}{\left((1 - V_f) K_m + V_f K_f + (3 K_f K_m / 4 \mu_m)\right)}$ |
| Shapery [39] | $\bar{\alpha} = \alpha_f + (\alpha_f - \alpha_m) \dfrac{K_m (K_f - \bar{K})}{\bar{K}(K_f - K_m)}$<br>Shapery upper bound (UB) is calculated taking the lower bound of Hashin-Shtrikman |

[40] estimate on the effective bulk modulus $\overline{K}$:

$$K_l = K_m + \frac{V_f}{\dfrac{1}{(K_f - K_m)} + \dfrac{(1 - V_f)}{(K_m + \mu_m)}}$$

In constrast the lower bound (LB) of the Schapery prediction is obtained with the upper Hashin-Shtrikman bound on the effective bulk modulus as

$$K_u = K_f + \frac{(1 - V_f)}{\dfrac{1}{(K_m - K_f)} + \dfrac{V_f}{(K_f + \mu_f)}}$$

where $K$ is the bulk modulus, $V$ denotes the volume fraction, $\mu$ is the shear modulus and subscripts $f$ and $m$ correspond to $ZrW_2O_8$ filler and HNBR matrix respectively; $u$ and $l$ are the upper and lower bounds.

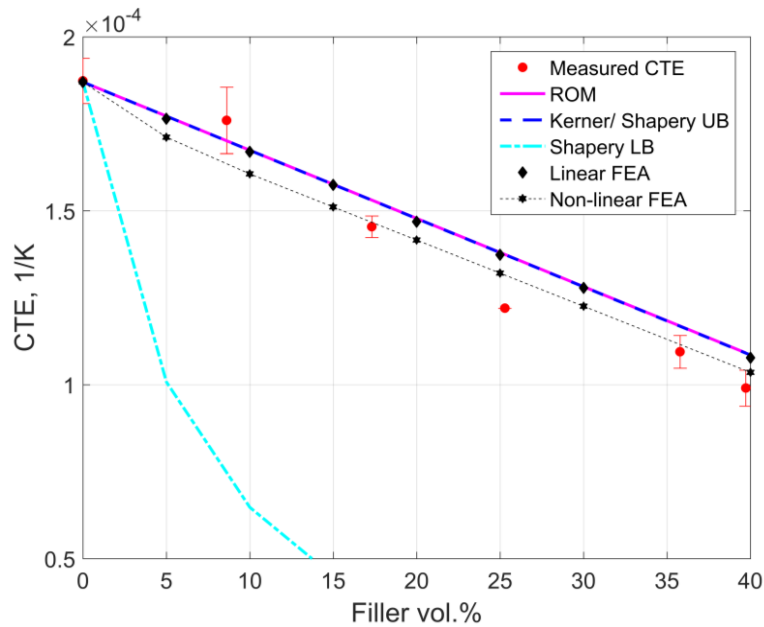

Figure 6. Linear CTE of the composites as a function of $ZrW_2O_8$ content at temperatures between 298 K and 403 K. Experimental points represent averages of 3 measurements (except for the 25.3 % filled HNBR) with error bars showing the standard deviation

The prediction models, except the lower bound of the Schapery model, capture the CTE reduction in the rubbery state reasonably well and can be used for engineering purposes. A similar relationship between the CTE and filler concentration was observed in polyurethane rubber compounds formulated with sodium chloride inclusions by Van Der Wal and coworkers [41] and also Holliday and Robinson [6] who demonstrated that the relationship for similar materials is linear and quite well described by the ROM. The estimates of the effective CTE provided by the ROM and linear FEA coincide and also match with predictions given in [27] who made use of the Levin formula [42] coupled with the Mori-Tanaka method [43] to get the effective mechanical characteristics and the thermal expansivity. These approaches give estimates that are close to the measured data. Nevertheless, slightly better solutions for the effective CTE can be offered by the non-linear FEA.

The adequacy of ROM predictions is questionable when the CTE of a glassy HNBR is concerned. This is demonstrated in Figure 7a, which graphically represents the relationship between the CTE and the $ZrW_2O_8$ content at temperatures below the glass transition of the HNBR compound (250 K), and in Figure 7b which presents the CTE results normalized to the value of CTE of the unfilled HNBR. The rate of the change of the thermal expansivity with increasing filling ratio is noticeably different for the composite material in the glassy state of the HNBR matrix. This effect is not captured by linear models like the ROM. A similar deficiency of the ROM was demonstrated in [23,25] on rigid thermosets with zirconium tungstate additions. In contrast to that, the linear FEA approach and the Kerner / Schapery UB models yield a more accurate estimation of the composite CTE. These predictions could be further improved if the exact Poisson's ratio or the bulk modulus of the glassy HNBR were known.

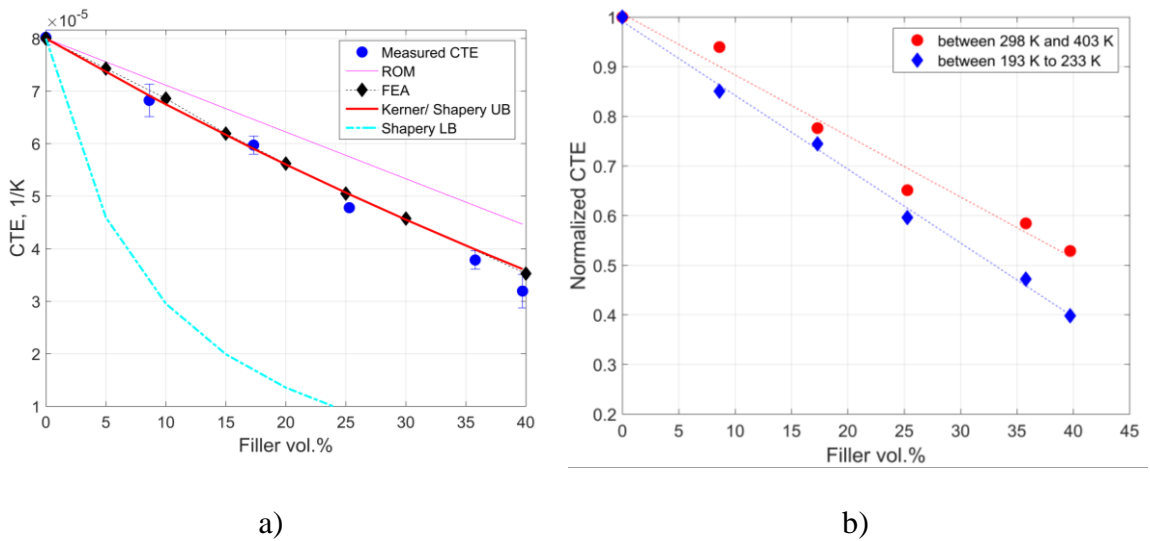

a)                      b)

Figure 7. Linear CTE of the composite as a function of $ZrW_2O_8$ content:
a) measured and predicted CTE at temperatures between 193 K to 233 K. The points and the error bars have the same significance as in Figure 5;

b) normalized CTE; straight lines represent a linear fit.

The drop of the normalized CTE due to the $ZrW_2O_8$ inclusions is more significant when HNBR is in the glassy state. Similar phenomena have been observed in styrene-butadiene rubber and natural rubber respectively, although they are attributed to the presence of carbon black [2] or variations of the cross-linking density [44]. The decrease in normalized CTE might be due to a substantial reduction of the free volume in the elastomer matrix inherent to the glassy state. As such, the polymer chains in HNBR are more densely packed, the material is stiffer and, thus, thermal displacements arising from the $ZrW_2O_8$ particles may be more efficiently transmitted over longer distances.

Kraus [2] asserts that this effect is caused by localized expansion of the free volume around carbon black particles due to thermal stresses arising from the CTE mismatch between them and the polymer matrix. This dilatation effect is believed [2] to superimpose on the normal thermal contraction in cooling resulting in a reduction of the CTE. The explanation is also transferable to the $ZrW_2O_8$ filler in rubber which should yield even higher thermal stresses. Indeed, FEA predicts that the biaxial stress in a thin layer of the glassy HNBR around the tungstate particle is substantial with the maximum equivalent stress reaching 7.2 MPa, as illustrated in Figure 8a, and cannot be relieved quickly as the relaxation processes in glassy polymers are to a large extent suppressed. Contrarily, in the rubbery state, the equivalent stress magnitude is low and it gets further reduced as the material undergoes stress relaxation by several mechanisms, which is not taken into account in the current simulation. Therefore, the stress effect contributes less to the CTE reduction at temperatures above the glass transition temperature of HNBR.

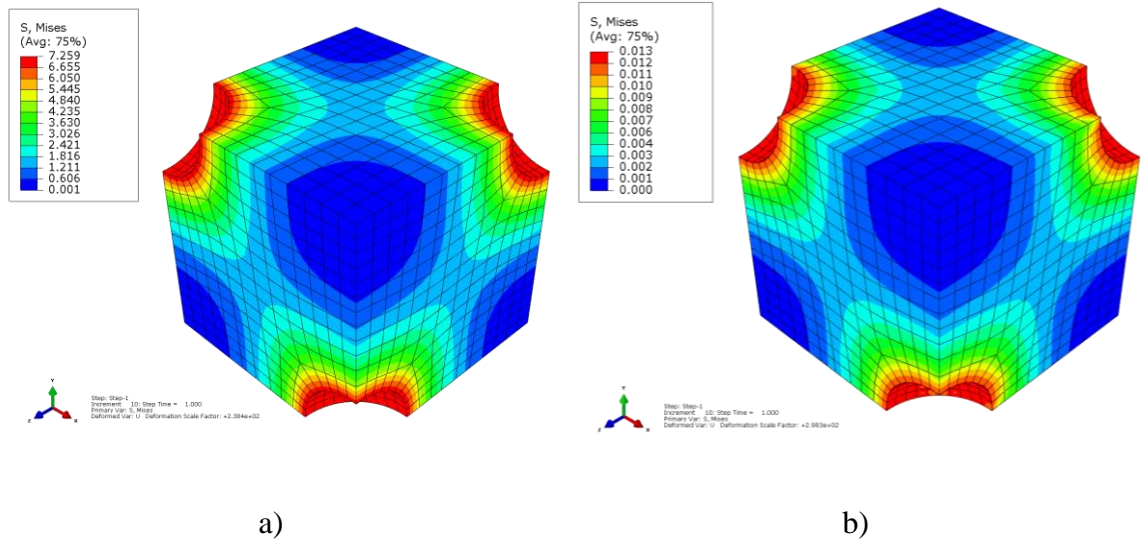

a)                  b)

Figure 8. Von Mises equivalent stress in a) glassy HNBR and b) rubbery HNBR matrix caused by the filler-matrix CTE mismatch. FEA simulation of HNBR filled with 10 % vol. fraction of $ZrW_2O_8$ subjected to 100 K temperature change

It should be noted that the predicted magnitude of the thermal stress in HNBR around the NTE inclusions are valid only if there is no particle detachment or voids in the interface. Conversely, such voids might have an impact on the thermal stresses and the CTE as well.

**3.3 Bulk Modulus Measurement**

The mechanical characteristics of the material is of much importance for the design of elastomer seals if high pressure and/or large deformations is required under operation conditions. The volumetric compression test results are now considered, while the outcome of the uniaxial extension test follows in the subsequent section.

The bulk modulus of the material is seen to rise with zirconium tungstate content as shown in Figure 9. It should be noted that the highest $ZrW_2O_8$ filled formulation (39.7 % volume fraction) reported before in this paper was not subjected to bulk modulus and other mechanical testing, due to a limited amount of this formulation being available.

The bulk modulus is increased by a factor of 1.45 in HNBR with 35,8 volume % of $ZrW_2O_8$, compared to the original rubber compound. The relationship between the bulk modulus and the $ZrW_2O_8$ concentration is in acceptable agreement with the theoretical prediction previously reported by Shubin *et al.* [27] considering the measurement error and potentially increasing porosity in the most highly filled compound. The authors [27] derived the following formula for calculating the composite bulk modulus $\overline{K}$ using the Mori-Tanaka method [43] which also matched with the Hashin-Shtrikman [40] lower bound:

$$\overline{K}(V_f) = K_m + \frac{V_f(K_f - K_m)(3K_m + 4\mu_m)}{3K_m + 4\mu_m + 3(1-V_f)(K_f - K_m)},$$

where the notation is the same as in Table 3. The values of the bulk modulus calculated using the equation above are also plotted in Figure 9.

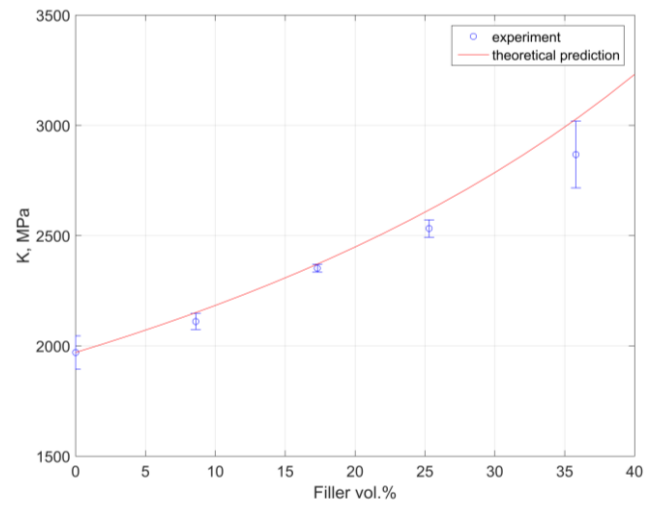

Figure 9. Bulk modulus of HNBR- $ZrW_2O_8$ composites as function of $ZrW_2O_8$ content. The graph points are averages of three samples with error bars showing the standard deviation.

## 3.4 Uniaxial tensile testing

The $ZrW_2O_8$ filled HNBR elastomer certainly becomes stiffer with increasing volume fraction of $ZrW_2O_8$. The effect is demonstrated by the initial portion of the stress-strain data in Figure 10 and the elastic modulus data in Figure 11. The elastic modulus is calculated by the linear regression of the stress-strain curve in a strain range of 0.05-0.25 %. The initial high slope in the stress-strain curves and the small-strain elastic modulus of the most highly filled HNBR apparently reflects a strong interaction between the rubber matrix and the $ZrW_2O_8$ inclusions. However, the tangent modulus becomes significantly lower at larger strains ≥10 % approaching the values typical for the unfilled rubber, which is below 10 MPa. This fall of the tangent modulus is presumably related to a gradual process of particle detachment from the rubber matrix and the accumulation of damage during the specimen extension. This appears to be supported by the SEM images shown in Figure 5.

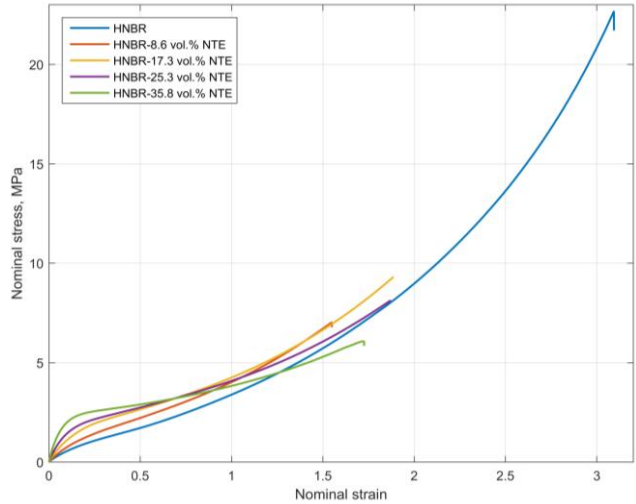

Figure 10. Representative stress-strain curves of the compounded materials measured at room temperature and at a strain rate of 0.083/sec

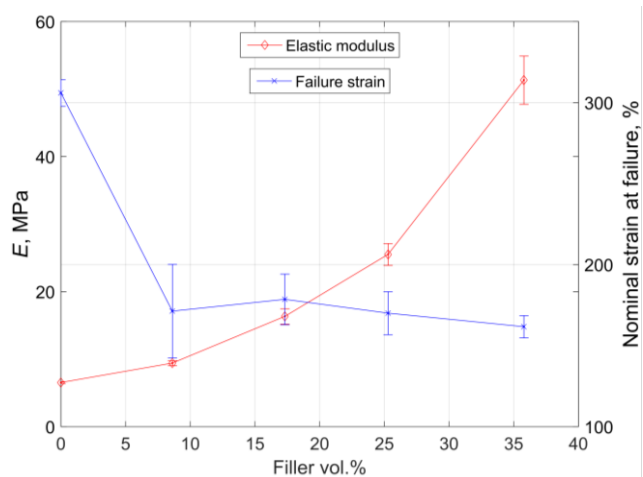

Figure 11. Tensile failure strain and elastic modulus as a function of $ZrW_2O_8$ content. The graph points are averages of five tests with error bars showing the standard deviation

As depicted in Figure 11, the presence of $ZrW_2O_8$ also negatively affects the total elongation which is decreased approximately 2 times for the composite with the lowest amount of the NTE filler in this study. It should be noted that these strains are still far beyond the strains typically occuring during seal applications (typically <50%), which are also not usually in uniaxial tension. Further additions >10% of the $ZrW_2O_8$ phase do not lead to a substantial decrease of the strain at failure of the composite material.

**3.5 Compression set**

Some correlation with the application in seals can be done through the compression set test data illustrated in Figure 12. It should be noted that the compression set test is not the best predictor of a sealing performance [45,46] of the tested material, however it may give a rough estimate of the potential for elastic recovery of the evaluated compounds and compare them. The test is also extensively used by the industry due to simplicity. The addition of zirconium tungstate to HNBR gives a slight increase in the compression set, from 12.5 % in the unfilled rubber to 15.8 % in the composite with 35.8 vol. % of $ZrW_2O_8$. These measured values of the compression set in the $ZrW_2O_8$ containing compounds can be considered acceptable and indicate general suitability for sealing applications, although further work would be required to confirm suitability for a particular seal.

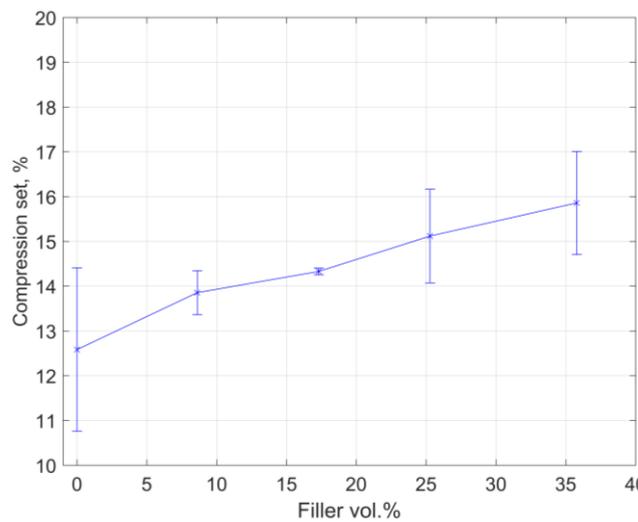

Figure 12. Compression set as a function of $ZrW_2O_8$ content. The graph points are averages of three samples in with error bars showing the standard deviation

**Conclusions**

Based on the experimental findings presented in this paper, the following conclusions can be made:

1) The HNBR-$ZrW_2O_8$ composites produced in this work exhibit a much lower coefficient of thermal expansion (CTE) than the unfilled HNBR compound. The peak CTE reduction of almost half of the original value in the unfilled HNBR is observed in the elastomer filled with approximately 40 % volume fraction of $ZrW_2O_8$. An even larger decrease in CTE to 0.4 of the initial CTE value is measured in the same compound, when it is cooled down below the glass transition of the HNBR.

2)      The correlation between the CTE and the $ZrW_2O_8$ volume fraction at temperatures above the glass transition temperature is reasonably well approximated by all of the analytical and numerical models considered in the work, except for lower bound of the Schapery model which gives an inadequate prediction. The rule of mixtures fails to provide a good estimate for the effective CTE in the glassy state. FEA as well as Kerner's expression and the upper bound of the Schapery model yielded much more accurate predictions of the composite CTE.

3)      The addition of $ZrW_2O_8$ to HNBR changes the mechanical characteristics of the material. The effect is observed in volumetric compression and uniaxial extension experiments yielding a higher initial slope of the stress-strain curves as compared to the unfilled HNBR. Like many inorganic fillers, the inclusion of $ZrW_2O_8$ leads to lower failure strains in the filled HNBR. Nevertheless, the lowest strain at failure values recorded are still well above the strains expected in sealing applications.

4)      The bulk modulus of the $ZrW_2O_8$ – HNBR composites was found to increase with $ZrW_2O_8$ content in agreement with previously reported theoretical predictions. The largest bulk modulus measured in the HNBR filled with 35.8 volume percent of the zirconium tungstate is about 1.45 times greater than that observed in the unfilled HNBR compound.

5)      The compression set increases slightly with the $ZrW_2O_8$ volume fraction, although this slight increase is unlikely to affect the use of these materials in seal applications.

**Acknowledgement**

This work is part of the collaborative project "Thermo Responsive Elastomer Composites for cold climate application" with the industrial partners FMC Kongsberg Subsea AS, STATOIL Petroleum AS, The Norwegian University of Science and Technology (NTNU) and the research institute SINTEF Materials and Chemistry. The authors would like to express their thanks for the financial support by The Research Council of Norway (Project 234115 in the Petromaks2 programme). The authors are also grateful to Huiting Jin at SINTEF Materials and Chemistry for the CTE measurements.

**Compliance with Ethical Standards**

**Conflict of Interest**

The authors declare that they have no conflict of interest.